\documentclass[11pt,twoside]{article}

\usepackage{asp2006}
\usepackage{epsf}
\usepackage{psfig}
\usepackage{lscape}

\usepackage{graphicx}
\usepackage{amssymb}

\begin{document}
\title{Flares from Sgr A* and their emission mechanism}   
\author{K.~Dodds-Eden\altaffilmark{1}, D.~Porquet\altaffilmark{2}, G.~Trap\altaffilmark{3,4}, E.~Quataert\altaffilmark{5}, 
S.~Gillessen\altaffilmark{1}, N.~Grosso\altaffilmark{2}, R.~Genzel\altaffilmark{1,6}, A.~Goldwurm\altaffilmark{3,4}, F.~Yusef-Zadeh\altaffilmark{7}, S.~Trippe\altaffilmark{8}, H.~Bartko\altaffilmark{1}, F.~Eisenhauer\altaffilmark{1}, T.~Ott\altaffilmark{1}, T.K.~Fritz\altaffilmark{1}, O.~Pfuhl\altaffilmark{1} }   

\begin{abstract} 
We summarize recent observations and modeling of the brightest Sgr~A* flare to be observed simultaneously in (near)-infrared and X-rays to date. Trying to explain the spectral characteristics of this flare through inverse Compton mechanisms implies physical parameters that are unrealistic for Sgr A*. Instead, a ``cooling break'' synchrotron model provides a more feasible explanation for the X-ray emission. In a magnetic field of about 5-30 Gauss the X-ray emitting electrons cool very quickly on the typical dynamical timescale while the NIR-emitting electrons cool more slowly. This produces a spectral break in the model between NIR and X-ray wavelengths that can explain the differences in the observed spectral indices.
\footnotetext[1]{Max-Planck-Institute for Extraterrestrial Physics, Garching, Germany\\
$^{}$ \hspace{-0.4cm} $^2$Observatoire astronomique de Strasbourg, Universit{\'e} de Strasbourg, CNRS, INSU, France\\
$^{}$ \hspace{-0.4cm} $^3$CEA, IRFU, Service d'Astrophysique, Centre de Saclay, France.\\
$^{}$ \hspace{-0.4cm} $^4$AstroParticule et Cosmologie (APC), Paris, France.\\
$^{}$ \hspace{-0.4cm} $^5$Department of Astronomy, University of California, Berkeley, USA\\
$^{}$ \hspace{-0.4cm} $^6$Department of Physics, University of California, Berkeley, USA\\
$^{}$ \hspace{-0.4cm} $^7$Department of Physics and Astronomy, Northwestern University, Evanston, USA\\
$^{}$ \hspace{-0.4cm} $^8$Institut de Radioastronomie Millim\'{e}trique, Saint Martin d'H\`{e}res, France}
\vspace{-0.9cm}
\end{abstract}
\section{Introduction} 
\vspace{-0.2cm}

X-ray flares have been observed from Sgr A* since 2001, and NIR flares since 2003 \citep{bag01,gen03}. How these flares are produced is still largely a mystery. Multiwavelength observations provide us with valuable information on the spectral properties of these flares \citep{eck04,eck06,eck09,yus06,yus08,mar08}. Sampling the flare SED at NIR and X-ray wavelengths where the emission may arise from different emission processes gives us clues as to the emission mechanisms involved and into the physical conditions in the region where the flare takes place. Investigating the emission mechanism is important not only because it gives us an insight into the physical conditions in the source region, but because it allows us to make the connection between NIR and X-ray wavelengths so that models for the time-variability at one wavelength can make predictions for the variability at the other. This gives us the potential to test and distinguish between flare models in ways that are not possible at one wavelength alone. 

\vspace{-0.2cm}
\section{Multiwavelength Results: the average SED}  \label{sec:multiwavelengthresults}
\vspace{-0.2cm}

\begin{figure}[!ht]
\begin{center}
\vspace{-3.0cm}
\hspace{-1.5cm}\includegraphics[width=6cm, clip = true, trim = 0 170 0 0]{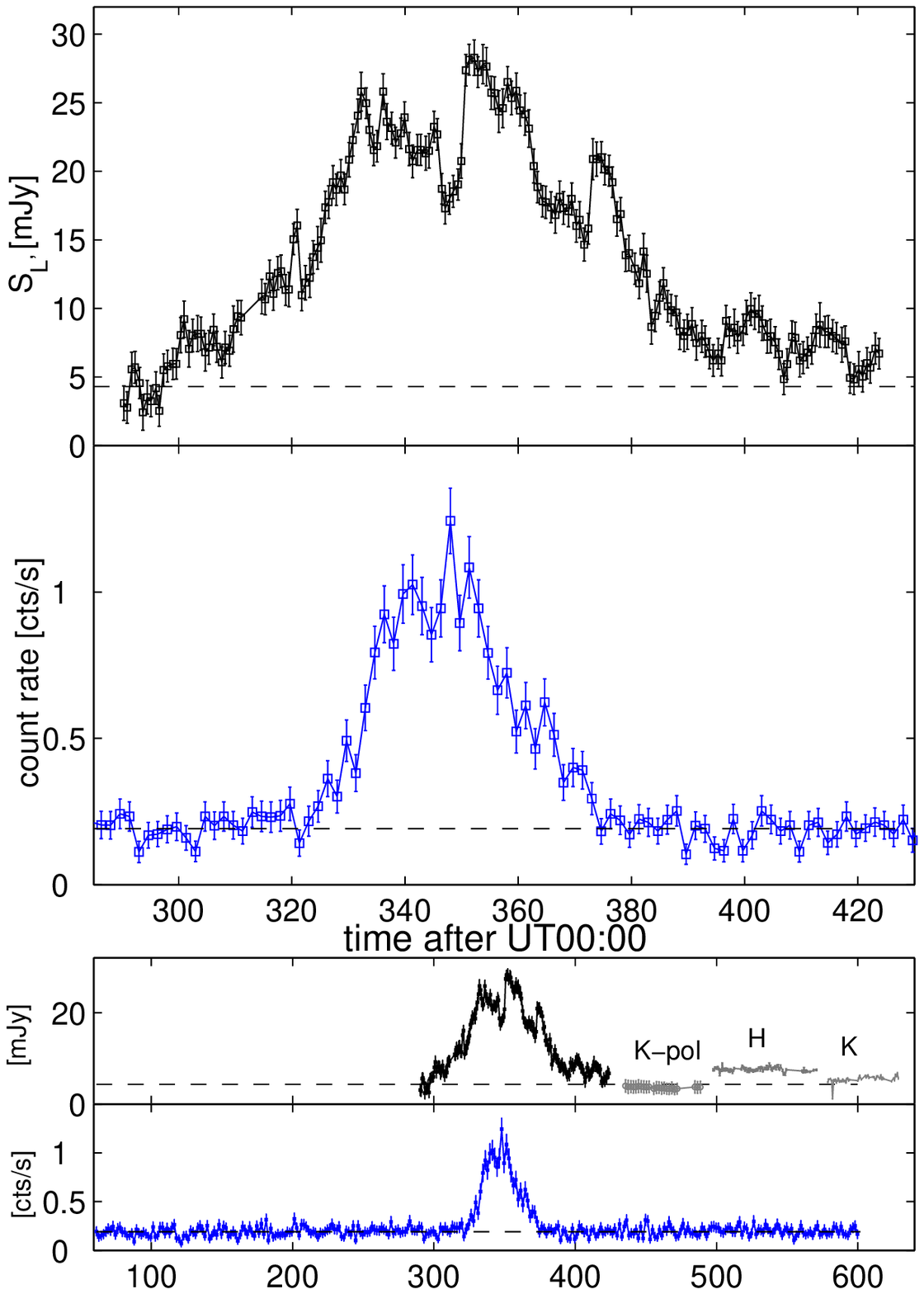}
\includegraphics[totalheight=9cm]{f5_rot.eps}
\vspace{-0.3cm}
\end{center}
\caption{\it (Left) IR (L'-band) and X-ray lightcurves from the April 4, 2007, and (right) the average spectral energy distribution showing in red the upper limit at 11.88$\mu$m, the average flux at 3.8$\mu$m and the unabsorbed X-ray spectrum assuming a power-law \citep[see ][ for more details]{dod09}.}\label{fig1}
\vspace{-0.1cm}
\end{figure}

Figure \ref{fig1} shows (left) simultaneous IR (L'-band) and X-ray lightcurves from a flare that was observed on April 4, 2007, and (right) the average spectral energy distribution \citep[for more details, see ][]{dod09,por08}. Both flares were bright, and together were by far the brightest flare that has ever been caught in a NIR/X-ray multiwavelength observation.

\noindent From MIR observations that were also simultaneous \citep{tra09}, no flare was detected, setting an upper limit on the 11.88 $\mu$m flux density of 57 mJy (dereddened using $A_{11.88\mu m}=1.7$). This upper limit implies that between 11.88$\mu$m and 3.8$\mu$m, the flare SED must have been blue in spectral energy ($\beta_{IR}>0$, where $\beta$ is defined as $\nu L_\nu \sim \nu^{\beta}$). The X-ray flare, on the other hand, was observed with a photon index $\Gamma = 2.3\pm0.3$ \citep[with 90\% confidence, ][]{por08} which implies the X-ray flare was soft in $\nu L_\nu$ ($\beta_X<0$). This April 4, 2007 X-ray flare was the second brightest X-ray flare yet observed; the brightest \citep{por03} also exhibited a soft ($\beta_X<0$) spectrum.

\vspace{-0.2cm}
\section{The emission mechanism} 
\vspace{-0.2cm} 

What do $\beta_{NIR}>0$ and $\beta_X<0$ imply for the emission mechanism behind the flares?
We investigated this in the context of three models for the origin of the X-ray radiation: 
\begin{enumerate}
\item \textbf{submm IC:} NIR flare = synchrotron, X-ray flare = submm photons inverse Compton scattered by NIR-emitting electrons (we assume the submm photons come from the known radio-submm source).
\item \textbf{SSC (synchrotron self Compton):} NIR flare = synchrotron, X-ray flare = NIR photons inverse Compton scattered by NIR-emitting electrons 
\item \textbf{cooling break synchrotron:} NIR flare = synchrotron, X-ray flare = synchrotron, with a cooling break between NIR and X-ray wavelengths. 
\end{enumerate}

We found it useful to approach this problem in a different way to previous investigations. Rather than using analytical power-law models for which, in order to apply the model, we must first assume the NIR and X-ray spectral indices are equal (an assumption which is not favoured by the data), we instead look at the problem from the point of view of where the peaks in the synchrotron and scattered spectra occur. We can make use of three well-known aspects of synchrotron radiation:
\begin{itemize}
 \item The synchrotron spectrum of a single electron with energy $\gamma$ peaks (in $\nu L_\nu$) at the critical frequency $\nu_c= 4.2\times10^6 \gamma^2 B$. \item An electron of energy $\gamma$ inverse Compton scatters a photon of frequency $\nu_{seed}$ up in frequency to $\nu_{IC} \approx \gamma^2\nu_{seed}$. \item The ratio of the inverse Compton luminosity to the synchrotron luminosity from a population of relativistic electrons follows ${L_{IC}}/{L_{synch}} = {U_B}/{U_{ph}}$.
\end{itemize}

As an extension of the first two points, populations of electrons with a characteristic energy $\gamma_*$ (a turnover in the underlying electron distribution, or a cutoff), corresponding to a synchrotron peak at $\nu_c(\gamma_*)$, will scatter a seed photon spectrum peaking at $\nu_{seed}$ to a frequency $\nu_{IC} = \gamma_*^2 \nu_{seed}$.

We can use these three properties together to make constraints on the physical parameters in the flaring region for the two inverse Compton scenarios  \citep{dod09}. Both scenarios are problematic: we find that in the submm IC case, the characteristic electron energy is constrained to be $\gamma_* \lesssim 1000$, the magnetic field $B\gtrsim24$ G and the size of the seed photon region $R < 0.1 R_S$, which is incompatible with the observed size of the submm IC region \citep[FWHM size $\approx 4 R_S$][]{doe08}. In the SSC case, we find $\gamma_* \lesssim 100$, $B\gtrsim2400$ G and a size of $R < 0.002 R_S$, which leads to an electron density $n_e>10^{10}$cm$^{-3}$. These values for the electron density and the magnetic field are two-three orders of magnitude higher than the electron densities and magnetic fields in the inner accretion flow \citep{yua03}.

The ``cooling break synchrotron'' model allows parameters more natural to the inner accretion flow. In this scenario the X-ray flare is also synchrotron radiation and there is a break in the synchrotron spectrum between IR and X-ray wavelengths due to the onset of dominant synchrotron cooling which causes a change in spectral index of $\Delta\beta=0.5$. For the kinds of magnetic fields expected in the inner regions of the accretion flow of $5$ to $30$G \citep{yua03}, X-ray emitting electrons cool fast on the dynamical timescale while NIR-emitting electrons cool slowly, so a cooling break between these wavelengths is very natural. Since both NIR and X-ray are optically thin synchrotron, we cannot obtain a constraint on the size of the flaring region as we did for the inverse Compton scenarios, except by requiring the \emph{absence} of an inverse Compton contribution to the X-ray luminosity (for reasonable parameters the inverse Compton spectra are blue in the X-ray range, and if they contributed the X-ray spectral index would not be red as observed). In a companion paper, \citet{tra09} have shown the IC/SSC components do not contribute for reasonable sizes and electron densities.

\vspace{-0.2cm}
\section{Conclusions}
\vspace{-0.2cm}
We have also carried out detailed numerical calculations which confirm the above constraints on the physical conditions in the region where the flare occurs \citep[][]{dod09}. The constraints we have outlined are of course not valid if the true spectral index of the NIR is $\beta_{NIR}<0$ rather than $\beta_{NIR}>0$. However, models where the synchrotron peak occurs at frequencies $\nu<\nu_{NIR}$ are not within the 90\% parameter confidence region when fitting our numerical models to the NIR and X-ray data (we fit the models and derived confidence regions using the X-ray spectral fitting package XSPEC). Furthermore, blue indices ($\beta_{NIR}>0$) for intermediate to bright emission from Sgr A* are favoured not just by our observational constraint on the MIR-NIR spectral index, but by evidence from other observations as well \citep{gen03,gil06,hor07}. 

There is much still to learn from the spectacular multiwavelength observation of April 4, 2007. The simultaneous lightcurves show intriguing differences: the L'-band lightcurve is broader (has longer duration) than the X-ray lightcurve, and it also has pronounced substructures while the X-ray lightcurve is comparatively smooth. These properties are not straightforward to understand in the context of any emission model. In the SSC scenario a shorter duration X-ray flare can arise naturally without changes in the magnetic field, due to the quadratic dependence of the SSC luminosity on the synchrotron luminosity, but there is no obvious solution to the substructure problem. For the cooling break synchrotron and the submm IC models fluctuations in magnetic field together with a general decrease in magnetic field during the flare could explain the substructure and duration problems, since in both cases the NIR emission is dependent on the magnetic field while the X-ray emission is not. Taking the spectral properties also into account, so far the cooling break synchrotron model appears to be the most promising model to explain the April 4, 2007 flare. Time dependent modeling of the lightcurves will give us further insights into the flare mechanism and the dynamics of their production. 



\end{document}